\documentclass[preprint,showpacs,preprintnumbers,amsmath,amssymb]{revtex4}

\def\undersim#1{\setbox9\hbox{${#1}$}{#1}\kern-\wd9\lower
    2.5pt \hbox{\lower\dp9\hbox to \wd9{\hss $_\sim$\hss}}}
\usepackage{amssymb}
\usepackage{amsmath}
\usepackage{graphicx}
\usepackage{color}

\def\undersim#1{\setbox9\hbox{${#1}$}{#1}\kern-\wd9\lower
    2.5pt \hbox{\lower\dp9\hbox to \wd9{\hss $_\sim$\hss}}}

\def\mv{{\mathbf v}}
\def\mr{{\mathbf r}}
\def\mK{{\mathbf K}}
\def\mr{{\mathbf r}}
\def\mq{{\mathbf q}}
\def\mk{{\mathbf p}}

\begin{document}

\title{Effect of relaxation time on the squeezed correlations of bosons for evolving
sources in relativistic heavy-ion collisions}

\author{{Wei-Ning Zhang$^{1}$\footnote{wnzhang@dlut.edu.cn}},
{Peng-Zhi Xu$^2$\footnote{xupengzhi@yeah.net}}}
\affiliation{$^1$School of Physics, Dalian University of Technology,
Dalian, Liaoning 116024, China\\
$^2$College of Computer Science and Technology, Jilin Normal University, Siping,
Jilin 136000, China}


\begin{abstract}
Squeezed back-to-back correlation (SBBC) of a boson-antiboson pair is sensitive to
the time distribution of the particle-emitting source, and the SBBC function for an
evolving source is expected to be affected by the relaxation time of the system.
In this paper, we investigate the effect of relaxation time on the SBBC function.
We put forth a method to calculate the SBBC function with relaxation-time modification
for evolving sources. The SBBC functions of $D^0\!{\bar D}^0$ in relativistic heavy-ion collisions are investigated by using a hydrodynamic model. It is found that the
relaxation time reduces the amplitudes of the SBBC functions.
This change becomes serious for a long
relaxation time and large initial relative deviations of the chaotic and squeezed
amplitudes from their equilibrium values, respectively, in the temporal steps. 

\end{abstract}


\maketitle

\section{Introduction}
In relativistic heavy-ion collisions, the interactions between the particles in
sources lead to a modification of boson mass in the sources, and thus give rise
to a squeezed correlation of boson-antiboson \cite{AsaCso96,AsaCsoGyu99}.
This squeezed correlation is caused by the Bogoliubov transformation between the
creation (annihilation) operators of the quasiparticles in the source and the free
observable particles, and impels the boson and antiboson to move in opposite
directions. So, it is also known as squeezed back-to-back correlation (SBBC) \cite{AsaCso96,AsaCsoGyu99,Padula06}. The measurements of the SBBC of bosons can
provide information on the interaction between the meson and the source medium,
and will be helpful to understand the properties of the particle-emitting sources
\cite{AsaCso96,AsaCsoGyu99,Padula06,Padula10,Zhang15,Zhang16}.

Hydrodynamics has been widely used in relativistic heavy-ion collisions to describe
the evolution of particle-emitting source. In hydrodynamic description it is assumed
that the source system is under the local equilibrium, i.e. the evolution is the
so-called ``quasi-static process". However, quasi-static process is a rough
approximation. Because the SBBC is sensitive to the time distribution of the source, dealing appropriately with the temporal factors is of interest in calculations of
the SBBC functions for evolving sources.

${\rm D}$ mesons contain a heavy quark (charm quark) produced during the early stage
of relativistic heavy-ion collisions. The SBBC of ${\rm D}$ mesons is stronger than
that of light mesons and useful for probing the source properties in the early stage
\cite{Zhang16,Padula10JPG,AGYang-CPL-18,XuZhang2019,XuZhang2019a}
In this paper, we put forth a method to calculate the SBBC function with relaxation-time
modification for evolving sources and investigate the effects of relaxation time on the
SBBC functions of ${\rm D}^0{\rm \bar D}^0$ in relativistic heavy-ion collisions using
hydrodynamic model VISH2$+$1~\cite{VISH2+1}.
We find that the SBBC functions decrease when relaxation time is taken into
consideration.
The change increases with increasing relaxation time and becomes serious for the large
relative deviations of the chaotic and squeezed amplitudes at the start of evolution 
from their equilibrium values, respectively.

The rest of this paper is organized as follows. In Sec. II, we present the formulas
of the SBBC functions for evolving sources with relaxation-time approximation. In
Sec. III, we investigate the influences of relaxation time on the SBBC functions of
${\rm D}^0{\rm \bar D}^0$ in relativistic heavy-ion collisions. Finally, a summary and
discussion are given in Sec. IV. \\

\section{Formulas}
The SBBC function of boson-antiboson with momenta $\mk_1$  and $\mk_2$ is defined as \cite{AsaCsoGyu99,Padula06}
\begin{equation}
\label{BBCf1}
C_{\rm SBB}(\mk_1,\mk_2) = 1 + \frac{\big|G_s(\mk_1,\mk_2)\big|^2}{G_c(\mk_1,\mk_1)
G_c(\mk_2,\mk_2)},
\end{equation}
where,
\begin{equation}
\label{Gc1}
G_c(\mk_1,\mk_2)=\sqrt{\omega_{\mk_1}\omega_{\mk_2}}\,\langle a^\dag(\mk_1) a(\mk_2)\rangle \equiv \sqrt{\omega_{\mk_1}\omega_{\mk_2}}\,\langle
g_c(\mk_1,\mk_2)\rangle,
\end{equation}
\begin{equation}
\label{Gs1}
G_s(\mk_1,\mk_2)=\sqrt{\omega_{\mk_1}\omega_{\mk_2}}\,\langle a(\mk_1) a(\mk_2)
\rangle \equiv \sqrt{\omega_{\mk_1}\omega_{\mk_2}}\,\langle g_s(\mk_1,\mk_2)\rangle,
\end{equation}
are the so-called chaotic and squeezed amplitudes, respectively \cite{AsaCsoGyu99,Padula06}, $\omega_\mk=\sqrt{\mk^2+m^2}$ is energy of free boson with
mass $m$, $a$ and $a^\dag$ are annihilation and creation operators of the free boson,
respectively, and $\langle\cdots\rangle$ indicates the ensemble average.

For a homogeneous thermal-equilibrium source with fixed volume $V$  and existing in
temporal interval [0--$\Delta t$] with time distribution $F(t)$, the amplitudes $G_c(\mk,\mk)$
and $G_s(\mk,-\mk)$ can be expressed as \cite{AsaCsoGyu99,Padula06}
\begin{eqnarray}
\label{Gc2}
G_c(\mk,\mk)=\frac{V}{(2\pi)^3}\,\omega_{\mk} \Big[|c_{\mk}|^2\,n_{\mk}+|s_{\mk}|^2
(n_{\mk}+1)\Big] \equiv \omega_{\mk} \langle g^0_c(\mk,\mk)\rangle,
\end{eqnarray}
\begin{eqnarray}
\label{Gs2}
G_s(\mk,-\mk)=\frac{V}{(2\pi)^3}\,\omega_{\mk} \Big[c_{\mk}\,s_{\mk}^*\,n_{\mk}
+c_{-\mk}s_{-\mk}^*\,(n_{-\mk}+1)\Big] \widetilde F(\omega_\mk,\Delta t)
\equiv \omega_{\mk} \langle g^0_s(\mk,-\mk)\rangle,
\end{eqnarray}
where
\begin{equation}
c^*_{\pm\mk}=c_{\pm\mk}=\cosh f_\mk, ~~~~~~
s^*_{\pm\mk}=s_{\pm\mk}=\sinh f_\mk, ~~~~~~
f_\mk =\frac{1}{2} \ln\left({\omega_\mk}/{\Omega_\mk}\right),
\end{equation}
where $\Omega_\mk =\sqrt{\mk^2 +m'^2}$ is energy of the quasi-particle with mass
$m'$ in source medium and $n_\mk$ is the boson distribution of the quasiparticle,
and $\widetilde F(\omega_\mk,\Delta t)$ is Fourier transform of time distribution.

For a evolving source, quantities $g_c(\mk,\mk)$ and $g_s(\mk,-\mk)$ may be given
by
\begin{equation}
\label{gct1}
g_c(\mk,\mk)=g^0_c(\mk,\mk)-\tau_c\, \frac{\partial g_c(\mk,\mk)}{\partial t},
\end{equation}
\begin{equation}
\label{gst1}
g_s(\mk,-\mk)=g^0_s(\mk,-\mk)-\tau_s\, \frac{\partial g_s(\mk,-\mk)}{\partial t},
\end{equation}
where $g^0_c(\mk,\mk)$ and $g^0_s(\mk,-\mk)$ are the quantities in equilibrium state,
given by Eqs.\,(\ref{Gc2}) and (\ref{Gs2}), and $\tau_c$ and $\tau_s$ are the
relaxation-time parameters related to the system ability of recovering equilibrium.
Relaxation-time approximation is an usual method for dealing with the quantities
in evolution systems. In this approximation, parameters $\tau_{c,s}$ are needed to
be smaller than the width of temporal interval, $\tau_{c,s}<\Delta t$, and they tend
to zero under the quasi-static condition.

Assuming $\tau_c=\tau_s={\bar \tau}$, Eqs.\,(\ref{gct1}) and (\ref{gst1}) give
\begin{equation}
\label{gct2}
g_c(\mk,\mk)=g^0_c(\mk,\mk)+\Delta^0_c e^{-t/{\bar \tau}} \approx g^0_c(\mk,\mk)
(1+\delta^0_c e^{-t/{\bar \tau}}),
\end{equation}
\begin{equation}
\label{gst2}
g_s(\mk,-\mk)=g^0_s(\mk,-\mk)+\Delta^0_s e^{-t/{\bar \tau}} \approx g^0_s(\mk,-\mk)
(1+\delta^0_s e^{-t/{\bar \tau}}),
\end{equation}
where $\Delta^0_{c,s}$ are differences of $g_{c,s}$ between the start of evolution 
and equilibrium, and it is assumed that the differences are approximately proportional
to $g^0_{c,s}$ with proportionality parameters $\delta^0_{c,s}$, respectively.
Parameter $\delta^0_c$ ($\delta^0_s$) denotes the relative deviation of the chaotic
(squeezed) amplitude at the start of evolution from its equilibrium value.
Then, the SBBC function $C_{\rm SBB}(\mk,-\mk)$ for an evolving source is given by,
\begin{equation}
\label{BBCf2}
C_{\rm SBB}(\mk,-\mk) = 1 + \frac{\Big| \langle g^0_s(\mk,-\mk)(1+\delta_s^0\,
e^{-t/{\bar \tau}})\rangle \Big|^2}
{\Big[\langle g^0_c(\mk,\mk)(1+\delta_c^0\,e^{-t/{\bar \tau}})\rangle\Big]
\Big[\langle g^0_c(-\mk,-\mk)(1+\delta_c^0\,e^{-t/{\bar \tau}}) \rangle\Big] }.
\end{equation}\vspace*{1mm}

For a hydrodynamic source, the chaotic and squeezed amplitudes, $G_c({\mk_1},{\mk_2})$
and $G_s({\mk_1}, {\mk_2})$, can be expressed in the relaxation-time approximation,
as \cite{AsaCsoGyu99,Padula06,Padula10,Zhang15,Zhang16},
\begin{eqnarray}
\label{Gchydro1}
&&G_c({\mk_1},{\mk_2})\!=\!\int\! \frac{d^4\sigma_{\mu}(r)}{(2\pi)^3}
K^\mu_{1,2}\, e^{iq_{1,2}\cdot r}\,\! \Big[|c'_{\mk'_1,\mk'_2}|^2
n'_{\mk'_1,\mk'_2}+\,|s'_{-\mk'_1,-\mk'_2}|^2(n'_{-\mk'_1,-\mk'_2}+1)\Big]
\nonumber\\
&&\hspace*{57mm} \times \Big[1+\delta^0_c\,e^{-t'/{\bar \tau}}\Big],
\end{eqnarray}
\begin{eqnarray}
\label{Gshydro1}
&&G_s({\mk_1},{\mk_2})\!=\!\int\! \frac{d^4\sigma_{\mu}(r)}{(2\pi)^3}
K^\mu_{1,2}\, e^{2 iK_{1,2}\cdot r}\!\Big[s'^*_{-\mk'_1,\mk'_2}
c'_{\mk'_2,-\mk'_1}n'_{-\mk'_1,\mk'_2}+c'_{\mk'_1,-\mk'_2} s'^*_{-\mk'_2,\mk'_1}
(n'_{\mk'_1,-\mk'_2} + 1) \Big]
\nonumber\\
&&\hspace*{59mm} \times \Big[1+\delta^0_s\,e^{-t'/{\bar \tau}}\Big],
\end{eqnarray}
where $d^4\sigma_{\mu}(r)=f_{\mu}(r)d^3\mr\,dt$ is the four-dimension element of
freeze-out hypersurface, $q^{\mu}_{1,2}=p^{\mu}_1-p^{\mu}_2$, $K^{\mu}_{1,2}=
(p^{\mu}_1+p^{\mu}_2)/2$, $c'_{\mk'_1,\mk'_2}$ and $s'_{\mk'_1, \mk'_2}$ are the
coefficients of Bogoliubov transformation between the creation (annihilation)
operators of the quasiparticles and the free particles, $n'_{\mk'_1,\mk'_2}$ is
the boson distribution associated with the particle pair in local frame, $\mk_i'~
(i=1,2)$ is local-frame momentum \cite{AsaCsoGyu99,Padula06,Padula10,Zhang15,Zhang16}.
In Eqs.\,(\ref{Gchydro1}) and (\ref{Gshydro1}), $[1+\delta^0_{c,s} e^{-t'/{\bar\tau}}]$ is the factor for relaxation-time influence, where $t'$ is local frame time.
Eqs.\,(\ref{Gchydro1}) and (\ref{Gshydro1}) will reduce to the usual forms as in
\cite{AsaCsoGyu99,Padula06,Padula10,Zhang15,Zhang16} when $\bar\tau=0$.

Dividing the whole time evolution into a series of time steps ($j=1,2,\cdots$) with
the same step width, we have
\begin{eqnarray}
\label{Gchydro2}
&&G_c({\mk_1},{\mk_2})\!=\sum_j \int\! \frac{f_{\mu}(r)d^3\mr}{(2\pi)^3}
K^\mu_{1,2}\, e^{-i\mq_{1,2}\cdot \mr}e^{iq^0_{1,2}t_j}\Big[|c'_{\mk'_1,\mk'_2}|^2
n'_{\mk'_1,\mk'_2}+|s'_{-\mk'_1,-\mk'_2}|^2(n'_{-\mk'_1,-\mk'_2}+1)\Big]
\nonumber\\
&&\hspace*{50mm} \times \Big[1+\delta^0_c \int^{\Delta\tau}_0 \!dt' D(t') e^{iq^0_{1,2}t(t')}e^{-t'/{\bar \tau}}\Big],
\end{eqnarray}
\begin{eqnarray}
\label{Gshydro2}
&&G_s({\mk_1},{\mk_2})\!=\sum_j \int\! \frac{f_{\mu}(r)d^3\mr}{(2\pi)^3}
K^\mu_{1,2}\, e^{-2i\mK_{1,2}\cdot \mr}e^{2iK^0_{1,2}t_j} \Big[s'^*_{-\mk'_1,
\mk'_2} c'_{\mk'_2,-\mk'_1}n'_{-\mk'_1,\mk'_2}+c'_{\mk'_1,-\mk'_2} s'^*_{-\mk'_2,
\mk'_1}
\nonumber\\
&&\hspace*{50mm} \times (n'_{\mk'_1,-\mk'_2} + 1) \Big]\Big[1+\delta^0_s \int^{\Delta\tau}_0 \!dt' D(t') e^{i2K^0_{1,2}t(t')}e^{-t'/{\bar \tau}}
\Big],
\end{eqnarray}
where $\Delta\tau$ is the step width of time in local frame, $t(t')=\gamma_v t'$,
$\gamma_v=1/\sqrt{1-\mv^2}$, $\mv$ is the velocity of fluid element, and $D(t')$
is local time distribution in each time step. Taking $D(t')$ to be an uniform
distribution, the relaxation-time factors for $G_c(\mk_i,\mk_i)~(i=1,2)$ and
$G_s(\mk_1,\mk_2)$ are,
\begin{equation}
\left[1+\delta^0_c\frac{\bar\tau}{\Delta\tau}\int^{\Delta\tau}_0\!\! dt'e^{-t'/{\bar
\tau}}\right] \approx \left[1+\delta^0_c\frac{\bar\tau}{\Delta\tau}\right], ~~~~~~
(\bar\tau <<\Delta\tau),
\end{equation}
\begin{equation}
\left[1+\delta^0_s\frac{\bar\tau}{\Delta\tau}\int^{\Delta\tau}_0\!\! dt'
e^{i2K^0_{1,2}t(t')} e^{-t'/{\bar\tau}}\right]\approx\left[1+\delta^0_c
\frac{\bar\tau}{\Delta\tau}\frac{1+i2K^0_{1,2}\gamma_v \bar\tau}
{1+(2K^0_{1,2}\gamma_v \bar\tau)^2} \right].
\end{equation} \\

\section{Results}
\begin{figure}[htbp]
\includegraphics[scale=0.85]{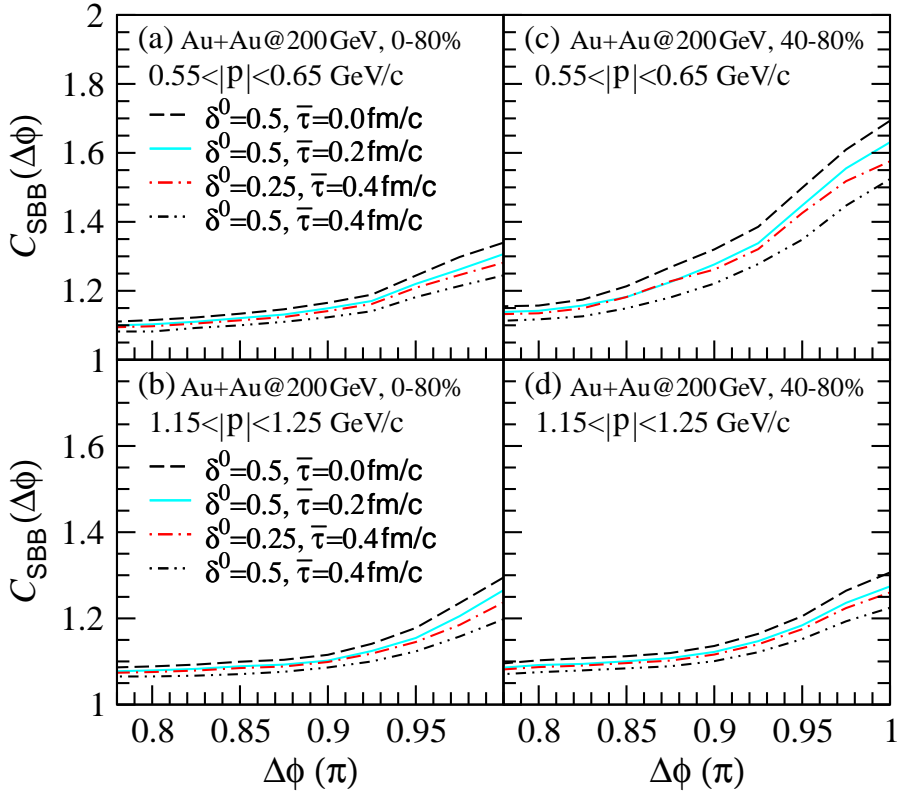}
\caption{The SBBC functions of ${\rm D}^0{\rm \bar D}^0$ pair in hydrodynamic
model for Au+Au collisions at $\sqrt{s_{NN}}=200$ GeV with 0--80\% and 40--80\% centralities and two momentum ranges. }
\label{DBBC-0848}
\end{figure}

We show in Figs.\,\ref{DBBC-0848}(a) and \ref{DBBC-0848}(b) the SBBC functions
$C(\Delta\phi)$ of the ${\rm D}^0 {\rm \bar D}^0$ pair in the hydrodynamic model
VISH2$+$1~\cite{VISH2+1} for Au+Au collisions at RHIC energy $\sqrt{s_{NN}}=200$ GeV
with 0--80\% centrality, and in momentum intervals 0.55--0.65\,GeV$/c$ and 1.15--1.25\,
GeV$/c$, respectively.
Here, $\Delta\phi$ is the azimuthal angle difference between the transverse momenta $\mk_{1T}$ and
$\mk_{2T}$ of the two D mesons, and the results of dashed, solid, dot-dashed, and
two-dot-dashed lines are for the parameters ($\delta^0_c=\delta^0_s=\delta^0=0.5$,
$\bar\tau=0$~fm$/c$, without relaxation-time modification), ($\delta^0_c=\delta^0_s=
\delta^0=0.5$, $\bar\tau=0.2$~fm$/c$), ($\delta^0_c=\delta^0_s=\delta^0=0.25$,
$\bar\tau=0.4$~fm$/c$), and ($\delta^0_c=\delta^0_s=\delta^0=0.5$,
$\bar\tau=0.4$~fm$/c$), respectively.
In the VISH2$+$1 model \cite{VISH2+1} used here, the  event-by-event initial conditions
of MC-Glb \cite{VISHb} are employed, the ratio of the shear
viscosity to entropy density of the QGP is taken to be 0.08 \cite{Shen11-prc,Qian16-prc}, and the freeze-out temperature is taken to be $T_f=150$ MeV according to the comparisons
of the model transverse momentum spectrum of $\rm D^0$ \cite{AGYang-CPL-18} with the
RHIC-STAR experimental data \cite{STAR-PRL113-2014}
(see the Fig. 3 in Ref. \cite{AGYang-CPL-18}).
In the calculations, the time step
width is taken to be 1~fm$/c$, free $\rm D^0$ meson mass is taken to be 1.865 GeV$/c^2$,
and the in-medium average mass and width are taken from the results of FMFK calculations
\cite{FMFK-PRC06,MFFK-PRL04,AGYang-CPL-18}.

\vspace*{5mm}

One can see from Figs.\,\ref{DBBC-0848}(a) and \ref{DBBC-0848}(b) that relaxation
time decreases the SBBC functions.
This change increases with increasing relaxation-time parameter $\bar\tau$ and the
parameter $\delta^0$. Relaxation-time parameter reflects the ability of recovering
equilibrium of system. In thermodynamics, high temperature and violent collisions
may increase the ability. The parameter $\delta^0$ is related to expanding velocity
of system. The SBBC functions in the lower momentum interval are higher than those
in the higher momentum interval because of a more serious in-medium mass modification
at low momenta than that at high momenta \cite{AGYang-CPL-18}, and the more serious oscillations of single-event SBBC functions at higher momentum \cite{Zhang15}
may also lead to a lower average SBBC function \cite{Zhang16,AGYang-CPL-18}.

We show in Figs.\,\ref{DBBC-0848}(c) and \ref{DBBC-0848}(d) the SBBC functions
$C(\Delta\phi)$ of the ${\rm D}^0 {\rm \bar D}^0$ pair in the hydrodynamic model
for Au+Au collisions at RHIC energy $\sqrt{s_{NN}}=200$ GeV with 40--80\% centrality,
and in momentum intervals 0.55--0.65\,GeV$/c$ and 1.15--1.25\, GeV$/c$, respectively.
One can see also that relaxation-time modification decreases the SBBC functions.
Similarly, the influence increases with increasing relaxation-time parameter $\bar\tau$
and the parameter $\delta^0$. Compared to the results for the collisions with 0--80\%
centrality, the SBBC functions for the collisions with 40--80\% centrality are higher.
The reason for this is mainly that the source temporal distribution is narrow in peripheral collisions \cite{Zhang16}. The contributions to the SBBC functions at lower
$\Delta\phi$ are mainly from the more peripheral collisions, which are small both in
spatial and temporal sizes \cite{AGYang-CPL-18}.
The differences between the SBBC functions for the collisions with 0--80\% and 40--80\% centralities become small in the higher momentum interval.
\\

\section{Summary and Discussion}
We have investigated the effects of relaxation time on the SBBC functions of
boson-antiboson pairs in relativistic heavy-ion collisions. A method for calculating
the SBBC functions for evolving sources with relaxation-time modefication is put forth.
Using the method in hydrodynamic model, we investigate the SBBC functions of ${\rm D}^0
{\rm \bar D}^0$ in Au+Au collisions at RHIC energy $\sqrt{s_{NN}}=200$ GeV with 0--80\%
and 40--80\% centralities. It is found that the relaxation time lead to a decrease of
the SBBC functions.
This change increases with increasing relaxation time and becomes serious for the large relative deviations of the chaotic and squeezed amplitudes at the start of evolution 
from their equilibrium values, respectively.

Relaxation-time approximation is an usual method to calculate quantities in a
near-equilibrium evolving system. In viscous hydrodynamic models, relaxation times
of shear and bulk viscosities are introduced, which may change the system space-time
structure. However, the relaxation times associated with the quantities are also needed
to be taken into account in their calculations with the hydrodynamic model, because
the quantities undergo a transition from nonequilibrium to equilibrium in each temporal
step. It is meaningful to consider appropriately relaxation time in calculating
a sensitive time-depend observable.

Using a hydrodynamic model, one can obtain the source temperature as a space-time
function. The final observed particles are assumed to be emitted thermally from a
four-dimension hypersurface with the fixed freeze-out temperature, which can be
determined by comparing the calculated observables, for example particle
transverse-momentum spectra, with experimental data.
In this paper, we use the viscous hydrodynamic model VISH2$+$1 \cite{VISH2+1} to
produce the freeze-out hypersurface and calculate the SBBC functions with and without
the relaxation-time term $\delta^0 e^{-t'/{\bar \tau}}$. The effect of relaxation
time reduces the SBBC functions. This effect may retain in an ideal hydrodynamic
model although the viscosity leads to a change of the freeze-out hypersurface.

Because the SBBC is caused by the particle mass modification in the source medium,
analyzing SBBC may perhaps become a new technique to extract the information of
the particle in-medium interactions in the future, although there are not experimental
data to compare with now.
On the other hand, it is hard to deal with the particle scattering in detail in a
bulk evolution model. In our model calculations we assume that the $\rm D$ mesons
have a mass shift and width, which are obtained from the FMFK calculations
\cite{FMFK-PRC06,MFFK-PRL04,AGYang-CPL-18}, in the sources because of the in-medium
interactions. More detailed studies of the influence of particle scattering on the
SBBC, based on a cascade model (for instance, a multi-phase transport model
\cite{Lin_PRC72_2005,HWangJHChen_NST32_2021}) or a hybrid model (for instance, the
hydro+UrQMD model \cite{SongBassHeinz_PRC83_2011,KarHuoPetBle_PRC91_2015}) will be
of interest. \\

\begin{acknowledgements}
This research was supported by the National Natural Science Foundation of
China under Grant Nos. 12175031 and 11675034.
\end{acknowledgements}


\begin{thebibliography}{99}

\bibitem{AsaCso96}
M. Asakawa and T. Cs\"org\H o, Heavy Ion Physics {\bf 4}, 233 (1996);
hep-ph/9612331.

\bibitem{AsaCsoGyu99}
M. Asakawa, T. Cs\"org\H o and M. Gyulassy, Phys. Rev. Lett. {\bf 83},
4013 (1999).

\bibitem{Padula06}
S. S. Padula, G. Krein, T. Cs\"org\H{o}, Y. Hama, P. K. Panda, Phys.
Rev. C {\bf 73}, 044906 (2006).

\bibitem{Padula10}
D. M. Dudek and S. S. Padula, Phys. Rev. C {\bf 82}, 034905 (2010).

\bibitem{Zhang15}
Y. Zhang, J. Yang, W. N. Zhang, Phys. Rev. C {\bf 92}, 024906 (2015).

\bibitem{Zhang16}
Y. Zhang and W. N. Zhang, Eur. Phys. J. C {\bf 76}, 419 (2016).

\bibitem{Padula10JPG}
S. S. Padula, D. M. Dudek, and Jr. O. Socolowski, J. Phys. G {\bf 37},
094056 (2010)

\bibitem{AGYang-CPL-18}
A. G. Yang, Y. Zhang, L. Cheng, H. Sun, and W. N. Zhang, Chin. Phys. Lett.
{\bf 35}, 052501 (2018).

\bibitem{XuZhang2019}
P. Z. Xu, W. N. Zhang, and Y. Zhang, Phys. Rev. C {\bf 99} 011902(R) (2019).

\bibitem{XuZhang2019a}
P. Z. Xu and W. N. Zhang, Phys. Rev. C {\bf 100} 014907 (2019).

\bibitem{VISH2+1}
H. Song and U. Heinz, Phys. Lett. B {\bf 658}, 279 (2008);
{\it ibid.}, Phys. Rev. C {\bf 77}, 064901 (2008).

\bibitem{VISHb}
C. Shen, Z. Qiu, H. Song {\it et al.}, arXiv:1409.8164;
https://u.osu.edu/vishnu/.

\bibitem{Shen11-prc}
C. Shen, U. Heinz, P. Huovinen, H. Song, Phys. Rev. C {\bf 84},
044903 (2011).

\bibitem{Qian16-prc}
J. Qian, U. Heinz, J. Liu, Phys. Rev. C {\bf 93}, 064901 (2016).

\bibitem{STAR-PRL113-2014}
L. Adamczyk {\it et al.} (STAR Collaboration), Phys. Rev. Lett. {\bf 113},
142301 (2014).

\bibitem{FMFK-PRC06}
C. Fuchs, B. V. Martemyanov, A. Faessler, and M. I. Krivoruchenko, Phys. Rev.
C {\bf 73}, 035204 (2006).

\bibitem{MFFK-PRL04}
B. V. Martemyanov, A. Faessler, C. Fuchs, and M. I. Krivoruchenko, Phys. Rev.
Lett. {\bf 93}, 052301 (2004).

\bibitem{Lin_PRC72_2005}
Z. W. Lin, C. M. Ko, B. A. Li, B. Zhang, and S. Pal, Phys. Rev. C {\bf 72},
064901 (2005).

\bibitem{HWangJHChen_NST32_2021}
H. Wang and J. H. Chen, Nucl. Sci. Tech. {\bf 32}, 2 (2021).

\bibitem{SongBassHeinz_PRC83_2011}
H. C. Song, S. A. Bass, and U. Heinz, Phys. Rev. C {\bf 83}, 024912 (2011).

\bibitem{KarHuoPetBle_PRC91_2015}
I. A. Karpenko, P. Huovinen, H. Petersen, and M. Hleicher, Phys. Rev. C
{\bf 91}, 064901 (2015).

\end{thebibliography}
\end{document}